\documentstyle[psbox]{WORKSHOP}
\begin{document}


\title{Revealing the nature of high-mass X-ray binaries\\through multi-wavelength and
statistical analyses} 


\author{
Arash Bodaghee,$^1$ John A. Tomsick,$^1$ Jerome Rodriguez $^2$\\[12pt]  
%
$^1$  Space Sciences Laboratory, University of California, Berkeley, USA \\
$^2$  Commissariat \`a l'Energie Atomique, Saclay, France \\
%
{\it E-mail(AB): bodaghee@ssl.berkeley.edu} 
}

\abst{We summarize the results of our long-running campaign to help understand the nature of high-mass X-ray binaries (HMXBs), emphasizing recent \emph{Suzaku} observations of IGR~J16207$-$5129 and IGR~J17391$-$3021. Thanks to the expanding ranks of HMXBs in our Galaxy, we are able to perform more reliable statistical analyses on the three currently-known sub-classes of HMXB: those with supergiant companions (SGXBs); those with Be companions (BEXBs); and the enigmatic Supergiant Fast X-ray Transients (SFXTs). We discuss new diagnostic tools, akin to the ``Corbet diagram,'' in which HMXBs tend to segregate based on their dominant accretion mechanism. We show how SFXTs span across the divided populations of BEXBs and SGXBs, bolstering the intriguing possibility that some SFXTs represent an evolutionary link. The use of HMXBs as tracers of recent massive star formation is revisited as we present the first ever spatial correlation function for HMXBs and OB star-forming complexes. Our results indicate that at distances less than a few kpc from a given HMXB, it is more likely to have neighbors that are known massive-star forming regions as opposed to objects drawn from random distributions. The characteristic scale of the correlation function holds valuable clues to HMXB evolutionary timescales.}

\kword{Gamma-rays: observations, catalogues -- X-rays: binaries}

\maketitle
\thispagestyle{empty}

\section{Multi-Wavelength Analyses}
High-mass X-ray binaries (HMXBs) feature a compact object such as a neutron star (NS) or a black hole candidate (BHC) accreting material from a young and massive donor star ($M > 10$\,$M_{\odot}$). A majority of HMXBs host main-sequence Be stars that have not filled their Roche lobe. These systems (BEXBs) are usually transient with flares produced whenever the sometimes wide and eccentric orbit brings the compact object close to its companion. Supergiant HMXBs (SGXBs) are accompanied by an evolved supergiant O or B star whose wind steadily feeds the compact object. Their emission tends to be persistent with low-level variability stemming from inhomogeneities in the wind. Supergiant Fast X-ray transients (SFXTs) are an emerging class of HMXB. They share properties from both established groups: like BEXBs, they are characterized by outbursts of short (hours) to long (days) duration whereby the peak intensity is several orders of magnitude greater than during quiescence; and like SGXBs, they are systems in which a compact object (usually a NS) is paired with a supergiant OB star. 

\begin{figure}[t]
\centering
\psbox[xsize=8cm]{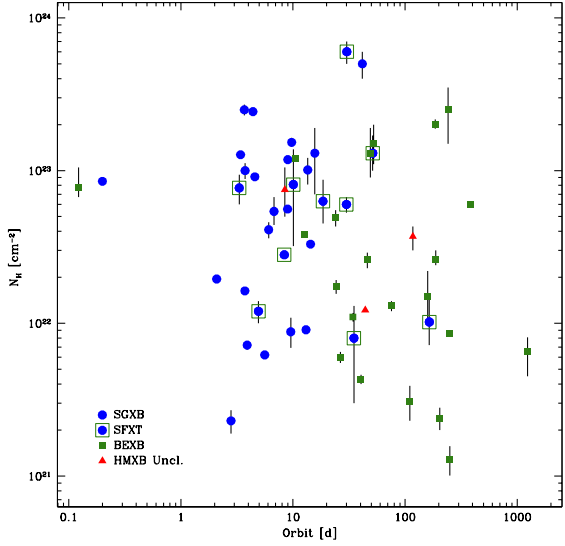}
\caption{Reported X-ray column densities for HMXBs as a function of orbital period are presented for SGXBs (blue disks, of which the SFXTs among them are boxed), BEXBs (green filled squares), and HMXBs with unclassified companions (red filled triangles).}
\label{nh_po}
\end{figure}

\begin{figure}[!t]
\centering
\psbox[xsize=8cm]{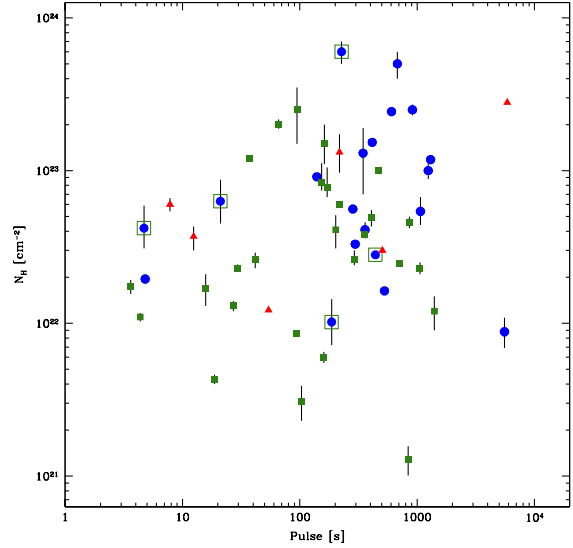}
\psbox[xsize=8cm]{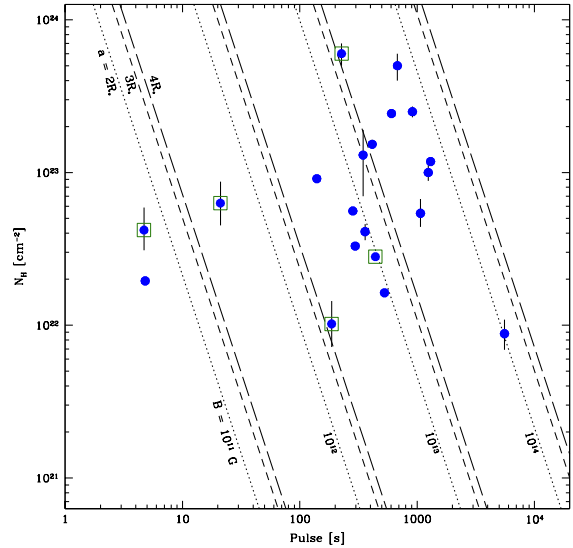}
\caption{Reported X-ray measured column densities for HMXBs as a function of pulsation period. Blue disks represent SGXBs (SFXTs are boxed), while green squares represent BEXBs. Unclassified HMXBs are denoted by red triangles. In the bottom panel, we focus on SGXBs and SFXTs and show equilibrium spin periods for a wind-accreting neutron star model from Waters \& van Kerkwijk (1989) assuming a variety of magnetic field strengths and binary separations. }
\label{nh_ps_po}
\end{figure}

While \emph{INTEGRAL} has proven adept at finding new HMXBs\footnote{An up-to-date list of \emph{INTEGRAL} sources and their properties can be found at: \texttt{http://irfu.cea.fr/Sap/IGR-Sources}}, in particular the heavily-absorbed or fast transient sources, its spatial resolution is of the order of an arcminute, i.e., too wide to permit a search for optical/IR counterparts in the busy fields in which HMXBs are located. It is only through dedicated follow-up observations with focusing X-ray telescopes such as \emph{Chandra}, \emph{Swift}-XRT, and \emph{XMM-Newton}, that an arc-second position can be obtained that is of use for optical/IR studies. Following up an IGR discovery with \emph{RXTE} and \emph{Suzaku} can provide more precise timing (\emph{RXTE}), broader spectral coverage (both), and a lower background (\emph{Suzaku}).

A number of research groups have put in place campaigns to systematically follow up new or previously-known HMXBs (and other source types): e.g., Chaty et al. (2008); Landi et al. (2010); Masetti et al. (2008--2010). Our group has been active in the analysis of all of the aforementioned telescopes: e.g., Bodaghee et al. (2010, 2011a); Rodriguez et al. (2009, 2010); Tomsick et al. (2008, 2009, 2011). Recently, we analyzed \emph{Suzaku} observations of the SGXB IGR~J16207$-$5129 and the SFXT IGR~J17391$-$3021 ($=$XTE J1739$-$302). The observation of IGR~J16207$-$5129 concludes with 30\,ks of emission consistent with no flux. There are no significant spectral changes, and the time scale is an order of magnitude too long for an off-state. We tentatively attribute this event to an eclipse of the X-ray emitter by its supergiant companion, but note that the demarcation between SFXTs and SGXBs is no longer as clear as was once believed, and so whatever mechanism is responsible for the quiescent state of SFXTs could be at play here. For IGR~J17391$-$3021, we observed quiescent emission interrupted by 3 weak flares in which the peak luminosity is only a factor of 5 times that of the quiescent phase. In this case, the change in luminosity was accompanied by a significant increase in the absorbing column indicating that the accretion of obscuring clumps is responsible for the weak flares. When placed in context of the long-term monitoring by \emph{Swift} (Romano et al. 2009), we find that these low-activity events (including those seen in this source by Bozzo et al. (2010) using \emph{XMM-Newton}) represent the most common state for this source ($60\pm5$\% of all observations in the 0.5--10\,keV band).

\begin{figure*}[t]
\centering
\psbox[xsize=17cm]{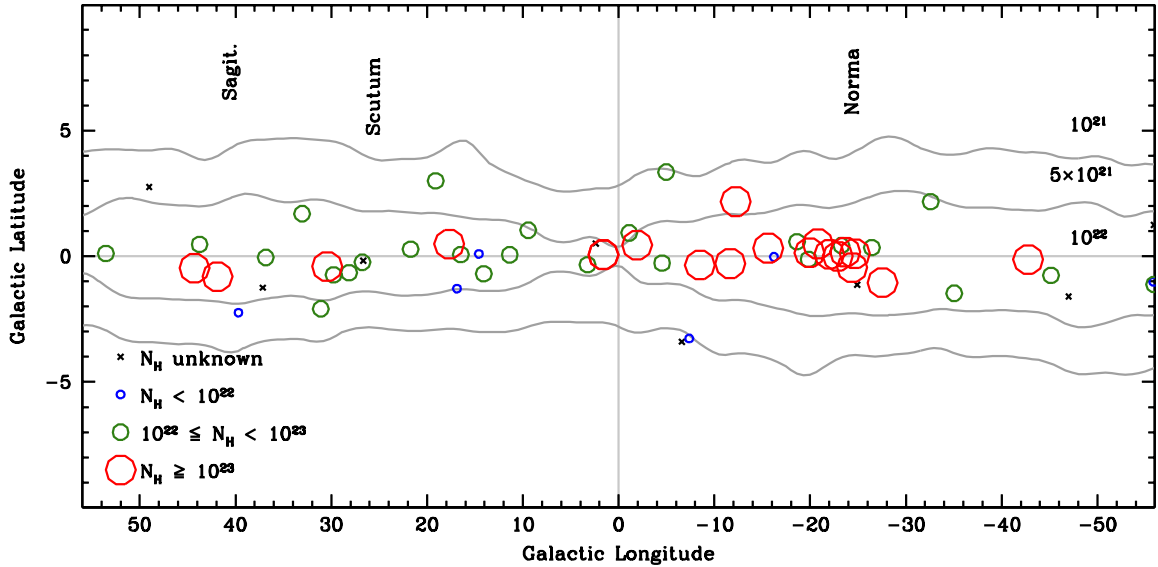}
\caption{The spatial distribution of HMXBs in Galactic coordinates with symbol size proportional to the X-ray measured column density as reported in the literature. The contours denote cumulative line-of-sight absorption levels of $10^{21}$, $5\times 10^{21}$, and $10^{22}$\,cm$^{-2}$ (Dickey et al. 1990). A cluster of heavily-obscured sources can be seen in the direction of the Norma Arm with no equivalent overdensity towards the other arms. }
\label{fov_disp_nh}
\end{figure*}

\section{Absorption \& Modulation}

The X-ray photoelectric absorption ($N_{\mathrm{H}}$) has been measured for a number of HMXBs. The intrinsic absorbing column is a direct measure of the amount of material in the vicinity of the compact object. As one would expect, the longer the binary period (i.e., the greater the average separation), the lower the $N_{\mathrm{H}}$, leading to a weak anti-correlation (Spearman rank correlation $R_{\mathrm{S}} = -0.3$, or a 2\% probability of being due to chance). Overall, SGXBs and BEXBs are segregated into distinct regions of the plot while the SFXTs bridge the divide. The vertical scatter in the plot is noticeably reduced when the column densities are normalized by the interstellar value expected along the line of sight (e.g., Dickey et al. 1990, Kalberla et al. 2005). 

Whether absorbing material is accreted or spun away, the pulsar spin frequency will be affected. When spin periods of the neutron stars in HMXBs are plotted as a function of the $N_{\mathrm{H}}$, we observe a segregation between BEXBs and SGXBs. The exceptions are SGXBs such as Cen X-3 that have atypical accretion regimes for their class. The SFXTs straddle the disparate groups. A weak correlation is detected with $R_{\mathrm{S}} = +0.3$ which corresponds to a 1\% probability of being due to chance. One explanation for the correlation is that the accretion of high-density wind material around the neutron star (with no preferred direction of angular momentum) will tend to slow the pulsar spin. Another explanation is that the perceived correlation is simply the result of two populations occupying different regions of the diagram. Once again, the vertical scatter can be diminished by normalizing the column density of each source with respect to the absorption expected in that direction.

\begin{figure*}[t]
\centering
\psbox[xsize=14.5cm]{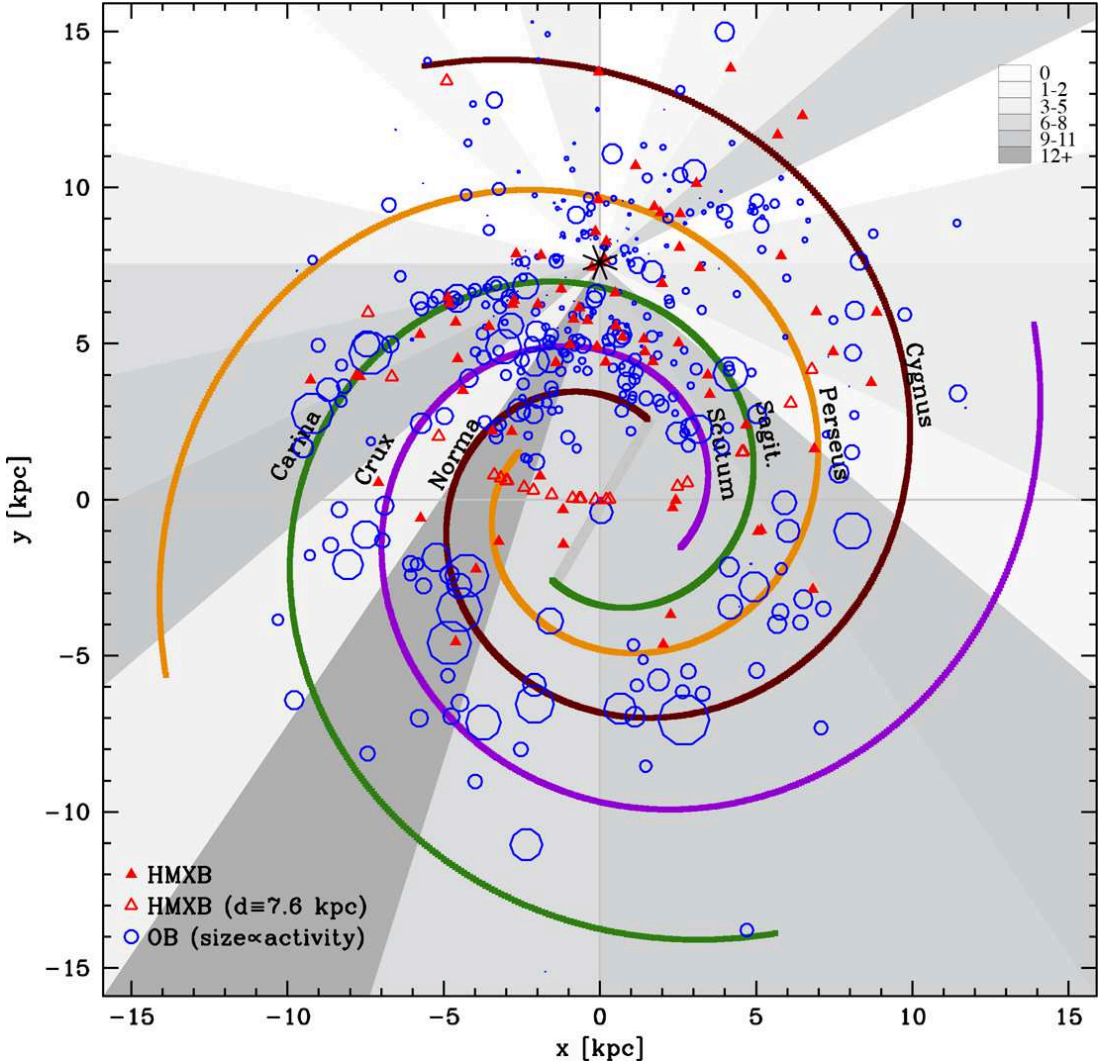}
\caption{Galactic distribution of HMXBs whose distances are known (75, filled triangles)
and the locations of OB star-forming complexes (464, circles) from Russeil (2003). The symbol size of the latter is proportional to the amount of activity in the complex. The spiral arm model of Vall\'ee (2008) is overlaid with the Sun situated at 7.6 kpc from the center. HMXBs whose distances are not known have been placed at 7.6 kpc (26, empty triangles). The shaded histogram represents the number of HMXBs in each 15$^{\circ}$ bin of galactic longitude.}
\end{figure*}

These figures lend support to the emerging model in which some SFXTs are descended from BEXBs and could therefore represent an evolutionary link between BEXBs and SGXBs (e.g., Negueruela et al. 2008, Chaty et al. 2011). First appearing in Bodaghee et al. (2007), these diagrams have been updated to distinguish the SFXT class, and to include the latest results such as MAXI~J1409$-$619, an unclassified HMXB with a spin period of $\sim$500\,s and $N_{\mathrm{H}} \sim 3\times10^{22}$\,cm$^{-2}$ (Kennea et al. 2010ab)

Figure\,\ref{nh_ps_po} (right panel) shows the spin period vs. $N_{\mathrm{H}}$ diagram with lines of equilibrium spin period as derived from the wind-accreting pulsar model of Waters et al. (1989). Canonical values are assumed for the mass of the neutron star and its supergiant companion, as well as for the supergiant wind velocity and the neutron star accretion rate. The column density for the model is obtained by integrating a range of wind densities along the line of sight. Several binary separation values and neutron star magnetic field strengths are considered. Pulsars in these HMXBs are spinning at periods that are longer than expected at equilibrium, unless the magnetic fields are stronger than $10^{12}$\,G (c.f., Santangelo et al. 2011). 

We continue to observe an asymmetry in the distribution of absorbed HMXBs in the inner spiral arms (Bodaghee et al. 2007). Figure\,\ref{fov_disp_nh} presents HMXBs in the central 150$^{\circ}$ of the Milky Way where the symbol size reflects the reported $N_{\mathrm{H}}$ value. Within 30$^{\circ}$ of the Galactic Center, there is a clear right-left asymmetry with 4 times as many absorbed sources at negative longitudes (right half of the figure) as there are at positive longitudes (left half of the figure). Absorbed sources are clustered in the direction of the Inner Perseus and Norma Arms with no equivalent overdensity towards the other arms. The Norma Arm happens to be the direction which features the highest number of massive-star forming regions (Russeil 2003). These are enormous clouds whose fragments spawn the massive precursor stars that eventually become HMXBs. In the next section, we will examine their spatial link in detail.

\section{Spatial Correlation}

\begin{figure*}[t]
\centering
\psbox[xsize=14cm]{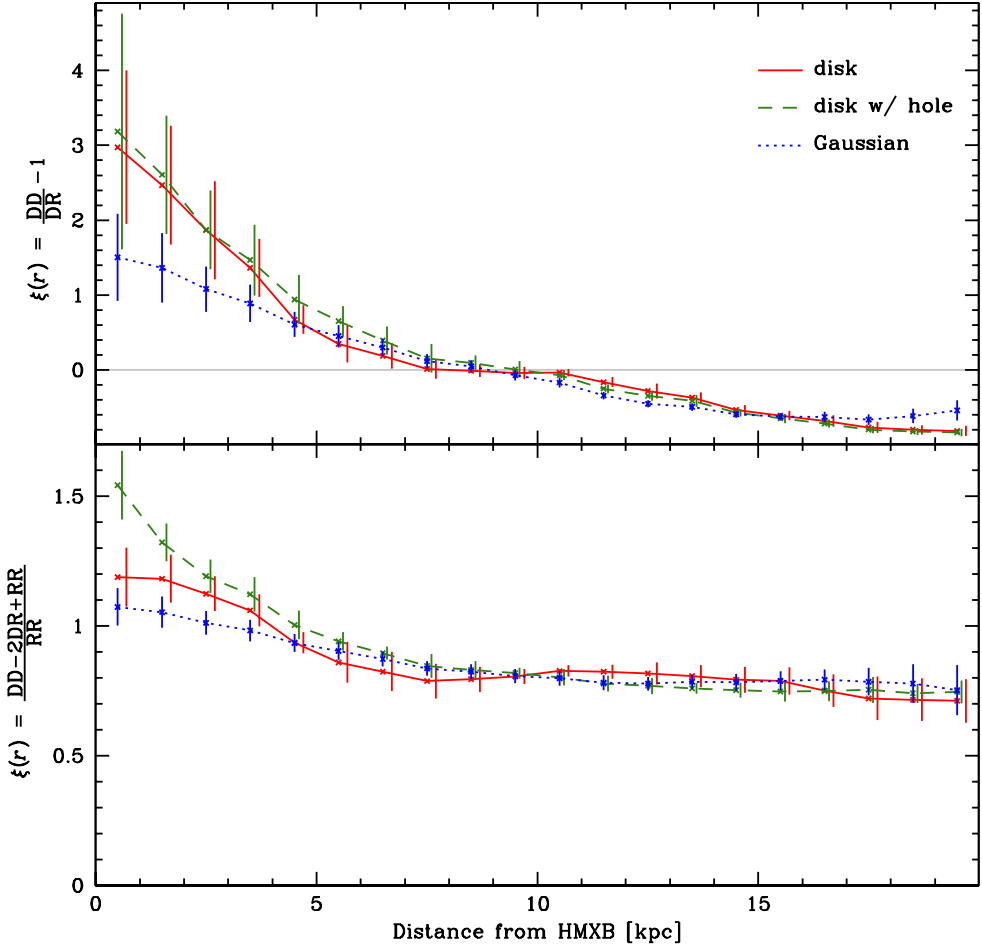}
\caption{Spatial correlation functions $\xi (r)$ from both estimators are presented for three randomization models: a Gaussian ring centered at 7.6\,kpc with $\sigma_{z} = 2$; a uniform disk for $0 < r < 16$\,kpc; and a uniform disk with a hole for $2 < r < 16$\,kpc. Error bars represent 3-$\sigma$ uncertainties. At short distances from a given HMXB, $\xi$ is greater than 1. This indicates that the closest neighbors of an HMXB tend to be observed OB associations rather than OB associations drawn from a random distribution.}
\label{fig_scf}
\end{figure*}

One of the goals of statistical analyses of HMXB populations is to compare their spatial distribution with that of the massive-star forming regions in which they are suspected of being born. Thus far, statistical compatibility between HMXBs and young star-forming systems has been demonstrated in distributions of galactic longitudes or galactocentric distance. The drawback is that these histograms consider only a single dimension. An entire axis is projected onto the other entailing information loss. This obstacle can be overcome by considering the spatial (or cross-) correlation function $\xi$. For a given HMXB in a volume element $\delta V_{1}$, the probability $\delta P$ (above Poisson) of finding an OB complex in a volume element $\delta V_{2}$ separated by a distance $r$ is:

\begin{equation}
\delta P = n_{1}n_{2}  \left [ 1 + \xi (r)  \right ] \delta V_{1}\delta V_{2}
\label{eq_prob}
\end{equation}

The mean number densities are given by $n_{1}$ and $n_{2}$ where the subscripts refer to the two populations: HMXBs (1) and OB complexes (2). Young objects such as HMXBs and OB stars are formed near the disk, i.e., we can safely assume that the displacement along the $z$-axis is negligible, so surface (rather than volume) elements are used. Around each HMXB, we construct rings of 1\,kpc thickness. In each ring, we count the number of observed OB complexes, and we keep a separate tally of the number of OB complexes from a random distribution. An HMXB-OB pair is referred to as a $DD_{12}$-pair (for data-data pair: the numbers designate the populations), while an HMXB-random OB pair is labelled $DR_{12}$ (for data-random). In this way, we construct $\xi$ for each radius according to the definition of Peebles (1980):

\begin{equation}
\xi (r) = \frac{ n_{R} DD_{12} }{ n_{D} DR_{12} } -1
\label{xi_pee80}
\end{equation}

Clearly, if $\xi = 0$, which implies that each ring contains as many DD-pairs as DR-pairs, then Eq.\,\ref{eq_prob} is simply a uniform Poissonian probability. However, if $\xi > 0$, then there is a higher chance of an HMXB having a neighbor that is an observed (rather than a randomized) OB complex. Landy \& Szalay (1993) propose a more reliable estimator for $\xi$ in which the variance is nearly Poissonian:

\begin{equation}
\xi (r) = \frac{DD_{12} - DR_{12} - DR_{21} + RR_{12}}{RR_{12}}
\label{eq_xi_pee80}
\end{equation}

Figure\,\ref{fig_scf} shows that the neighbor of a randomly-selected HMXB is more likely to be a known OB complex than expected from Poisson statistics. For $r \leq 3$\,kpc, the clustering signal from the Peebles (1980) estimator is 7--11$\sigma$ in excess of Poisson (up to 17$\sigma$ for the Landy \& Szalay (1993) estimator). The observed and random surface-density distributions are statistically compatible only at large radii (i.e. $r > 5$\,kpc).

In other words, HMXBs and OB associations are clustered together, as one would expect since HMXBs are young systems that have not ventured far from their likely birthplaces. This is the first time the relationship has been demonstrated in Cartesian space in our Galaxy. The characteristic scale of $\xi$ holds clues to the average offset of the HMXB population compared to the current crop of star-forming regions. In turn, this will help constrain the amount of migration and kinematical timescales of HMXBs.

It is reasonable to expect symmetry in the distribution of HMXBs and OB complexes around the galactic disk and our randomization models assumed this. However, the fact that there are less HMXBs detected (with distances measured) behind the Bulge will surely add bias to $\xi$. Therefore, we generated $\xi$ for a restricted boundary corresponding to a circle of radius 8 kpc around the Sun---a perimeter inside of which most HMXBs and OB complexes should be detectable and their distances known with reasonable accuracy. When only pair counts of objects (53 HMXBs and 363 OB complexes) in the Solar neighborhood ($r  < 8$\,kpc) were considered, the clustering signal at small radii persisted with a statistical significance of 7$\sigma$ for $r  \leq 1$\,kpc.

Clustering between HMXBs and OB complexes is expected. How would $\xi$ react if HMXBs were compared to a population of sources for which no spatial correlation is anticipated? We tested the correlation functions of HMXBs against a set of 133 globular clusters from Bica et al. (2006) that are located less than 20\,kpc from the GC. Globular clusters contain older populations such as cool KM dwarf stars and low-mass X-ray binaries (LMXBs). Unlike HMXBs, globular clusters (and LMXBs) are densely packed in the Galactic Bulge and their numbers drop exponentially with increasing radius from the GC. The random distribution of globular clusters was thus modeled as an exponential decay law adjusted to be consistent with the observed distribution. The $\xi$ between HMXBs and globular clusters is consistent with 0 for $r \le 1$\,kpc from an HMXB. In other words, for any given HMXB, its immediate neighbors were just as likely to be drawn from the observed set as from the random set. Shot noise dominates at large distances. 

Currently, we are working on accounting for other observational biases, and we are examining the behavior of $\xi$ under different evolutionary timescales. Further discussions of these results will soon be submitted for publication as Bodaghee et al. (2011b).

\section{Conclusion}

Multi-wavelength analysis of an HMXB can be used to reveal the mechanisms responsible for the emission, describe the composition and morphology of material surrounding (and in some cases enshrouding) the system, and permit a firm identification of the optical/IR counterpart. Multi-variate analysis of the spatial, timing, and spectral parameters of HMXB populations has served to confirm theoretical predictions (such as the link between HMXBs and OB associations), and to uncover trends that await explanation (such as the correlation in the $N_{\mathrm{H}}$ vs. spin period diagram). With large and growing samples representing each of the three known HMXB classes, we can achieve better statistics while exploring the parameter space occupied by HMXBs. A synthetic study of the different populations of HMXB will help reveal clues to the nature and evolutionary history of each class, which will help position each group within a unified model of HMXBs.

\section*{References}

\re
Bica E. et al., 2006, A\&A, 450, 105

\re
Bodaghee A. et al., 2007, A\&A, 467, 585

\re
Bodaghee A. et al., 2010, ApJ, 719, 451

\re
Bodaghee A. et al., 2011a, ApJ, 727, 59

\re
Bodaghee A. et al., 2011b, in prep.

\re
Bozzo E. et al., 2010, A\&A, 519, 6

\re
Chaty S. et al., 2008, A\&A, 484, 783

\re
Chaty S. et al., 2011, ESA-SP, submitted

\re
Dickey J.M. et al., 1990, ARA\&A, 28, 215

\re
Kalberla P.M.W. et al., 2005, A\&A, 440, 775

\re
Kennea J.A. et al., 2010a, ATel, 2962, 1

\re
Kennea J.A. et al., 2010b, ATel, 3060, 1

\re
Landi R. et al., 2010, MNRAS, 403, 945

\re
Landy S.D. \& Szalay A.S., 1993, ApJ, 412, 64

\re
Liu Q.Z. et al., 2006, A\&A, 455, 1165

\re
Masetti N. et al., 2008, A\&A, 482, 113

\re
Masetti N. et al., 2009, A\&A, 495, 121

\re
Masetti N. et al., 2010, A\&A, 519, 96

\re
Negueruela I. et al., 2008, AIPC, 1010, 252

\re
Peebles P.J.E., 1980, ``The Large Scale Structure of the Universe''

\re
Rodriguez J. et al., 2009, A\&A, 508, 889

\re
Rodriguez J. et al., 2010, A\&A, 517, 14

\re
Romano P. et al., 2009, MNRAS, 399, 2021

\re
Russeil J., 2003, A\&A, 397, 133

\re
Santangelo A. et al., 2011, these proceedings

\re
Tomsick J.A. et al., 2008, ApJ, 685, 1143

\re
Tomsick J.A. et al., 2009, ApJ, 701, 811

\re
Tomsick J.A. et al., 2011, ApJ, 728, 86

\re
Vall\'ee J.P., 2008, ApJ, 681, 303

\re
Waters L.B.F.M. \& van Kerkwijk M.H., 1989, A\&A, 223, 196

\label{last}

\end{document}